\documentclass[twocolumn,amsmath,amssymb,superscriptaddress,prl]{revtex4-1}

\usepackage{graphicx}
\usepackage{dcolumn}
\usepackage{bm}
\usepackage{color}
\def\Vec#1{\bm{#1}}

\begin{document}


\title{
Self-learning Hybrid Monte Carlo: A First-principles Approach
}

\author{Yuki Nagai}
\email{nagai.yuki@jaea.go.jp }
\affiliation{CCSE, Japan  Atomic Energy Agency, 178-4-4, Wakashiba, Kashiwa, Chiba, 277-0871, Japan}
\affiliation{
Mathematical Science Team, RIKEN Center for Advanced Intelligence Project (AIP), 1-4-1 Nihonbashi, Chuo-ku, Tokyo 103-0027, Japan
}

\author{Masahiro Okumura}
\affiliation{CCSE, Japan  Atomic Energy Agency, 178-4-4, Wakashiba, Kashiwa, Chiba, 277-0871, Japan}

\author{Keita Kobayashi}
\affiliation{Research Organization for Information Science and Technology (RIST), 2-4, Shirakata, Tokai-mura, Ibaraki 319-1106, Japan}

\author{Motoyuki Shiga}
\email{shiga.motoyuki@jaea.go.jp }
\affiliation{CCSE, Japan  Atomic Energy Agency, 178-4-4, Wakashiba, Kashiwa, Chiba, 277-0871, Japan}

\date{\today}

\begin{abstract}

We propose a novel approach called Self-Learning Hybrid Monte Carlo (SLHMC) which is a general method to make use of machine learning potentials to accelerate the statistical sampling of first-principles density-functional-theory (DFT) simulations.
The trajectories are generated on an approximate machine learning (ML) potential energy surface. The trajectories are then accepted or rejected by the Metropolis algorithm based on DFT energies.
In this way the statistical ensemble is sampled exactly at the DFT level for a given thermodynamic condition.
Meanwhile the ML potential is improved on the fly by training to enhance the sampling, whereby the training data set, which is sampled from the exact ensemble, is created automatically.
Using the examples of $\alpha$-quartz crystal SiO$_2^{}$ and phonon-mediated unconventional superconductor YNi$_2^{}$B$_2^{}$C systems, we show that SLHMC with artificial neural networks (ANN) is capable of very efficient sampling, while at the same time enabling the optimization of the ANN potential to within meV/atom accuracy.
The ANN potential thus obtained is transferable to ANN molecular dynamics simulations to explore dynamics as well as thermodynamics. This makes the SLHMC approach widely applicable for studies on materials in physics and chemistry.

\end{abstract}

\maketitle

\paragraph{Introduction.}
First-principles molecular dynamics based on density functional theory (DFT-MD) is a powerful tool to simulate a variety of materials in physics and chemistry \cite{Marx}. 
However, reducing the computational effort required for DFT-MD remains a key issue for its broader application to phenomena on large length- and time-scales. 
The use of artificial neural networks (ANN), which imitate
DFT energies by machine learning, is seen as a promising solution to this issue \cite{Blank1995,Brown1996,Lorenz2004,Behler2007}. 
The branch of the research about machine learning molecular simulations has grown rapidly in the last decade \cite{Behler2015,Artrith2011,Artrith2017,ZLi2015,Artrith2014,Li,Liarxiv,Jacobsen,Artrith2018,Jinnouchi2019,Bartok2010,Smith2017,Imbalzano} after an influential paper by Behler and Parrinello \cite{Behler2007}  
laid down a general framework of setting up and training ANN potentials from
DFT data sets, and running ANN-MD simulations for condensed matter systems.

\begin{figure}[t]
\begin{center}
     \begin{tabular}{p{ 1 \columnwidth}} 
           \resizebox{1 \columnwidth}{!}{\includegraphics{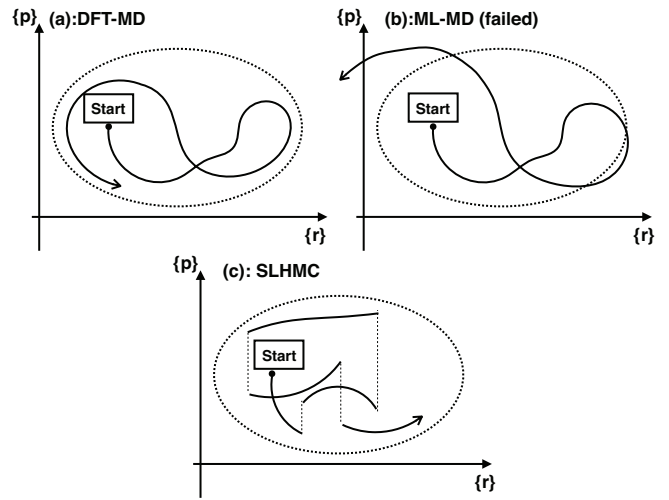}} 
    \end{tabular}
\end{center}
\caption{
\label{fig:fig1} 
Schematic figure of trajectories in phase space. (a) Trajectories of DFT-MD, (b) ML-MD, and (c) SLHMC. If the ML potential is not well-trained, the ML-MD trajectory might fall outside the range of the DFT ensemble (depicted by a dotted oval), while the accepted SLHMC trajectory always stays inside.
}
\end{figure}

The training of machine learning (ML) potentials must be based on sufficient amounts of DFT-derived results to cover all the configuration space, which corresponds to a statistical ensemble in the case of systems in thermal equilibrium. 
Usually the training sets for ML potentials are chosen before starting ML-MD simulations. 
Many useful methods, such as generic algorithms \cite{Artrith2018} and CUR decompositions \cite{Imbalzano} etc., were suggested for properly choosing training sets. 
If sufficient care is not taken, it is possible that ML-MD simulations may break down suddenly when the trajectory finds its way into an uncovered part of the phase space, see Fig.~\ref{fig:fig1}. 
Therefore it would be beneficial to establish a way to somehow cover the ensemble space in an automatic manner.

Here we propose a novel approach for training ML potentials based on the exact statistical ensemble at a given thermodynamic condition, e.g.,
isothermal, isobaric ensembles, using the hybrid Monte Carlo technique \cite{Gottlieb,Duane,Mehlig}. 
This allows one to access exact results out of approximate ML potentials, and, at the same time, create an objective and unique data set for ML training. 
Importantly, we want to circumvent
the use of costly DFT-MD computations as much as possible, which can be achieved by training the ML potential on systems with small sizes and short time scales.
Our idea is to use the dual-level HMC method \cite{Iftimie2000,Gelb,Nakayama,Suzuki}. The ensemble is sampled by generating trial moves of the trajectory from an approximate ML potential energy surface, which are then either accepted or rejected by performing exact
DFT energy calculations at long time step intervals.
Note that the ensemble created is theoretically exact irrespective of the quality of the approximate ML potentials. 
Its efficiency is strongly dependent, however, on the quality of the ML potential, as this affects acceptance ratio. 
It will be shown herein that the acceptance ratio tends to improve as the ML potential is trained and
updated iteratively,
which is done automatically on the fly during the computations. 
Note that the ML potential is used here in an auxiliary manner to produce exact ensembles, which distinguishes our concept from previous works where the ML potential was used as an approximation
 to imitate exact DFT potentials.


We call our method the self-learning hybrid Monte Carlo (SLHMC) method, since it is akin to the self-learning Monte Carlo (SLMC) method \cite{Liu}.
The SLMC method was recently introduced in the field of many-electron systems to speed up MC simulations by using efficient global updates informed by machine learning techniques \cite{Liu,LiuShen,Xu,Chen,Nagai,NagaiBPNN,HuitaoNN}. 
SLHMC extends the idea of SLMC to equation-of-motion based moves to enable efficient global updates for atomistic and molecular simulations.

This letter is organized as follows:
We first explain the basic theory and computational procedure of SLHMC. 
We use example calculations on SiO$_2^{}$ ($\alpha$-quartz) to demonstrate the accuracy and efficiency of the method.
Calculations for the phonon-mediated superconductor YNi$_2^{}$B$_2^{}$C are then used to demonstrate the ability to construct accurate ML potentials such that the ML-MD simulations are stable at long times.

\paragraph{Self-learning Hybrid Monte Carlo Method.}
A trial move uses
 Hamilton's equations of motion derived from the ML potential energy surface, $V_{\rm ML}^{}$,
\begin{equation}
\dot{\Vec{p}}_i^{} = - \frac{\partial V_{\rm ML}^{}(\{\Vec{r} \},t)}{ \partial \Vec{r}_i^{}},
 \ \ \ \ \ \
\dot{\Vec{r}}_i^{} = \frac{\Vec{p}_i^{}}{m_i^{}}, \label{eq:ri}
\end{equation}
where $\Vec{p}_i^{}$, $\Vec{r}_i^{}$ and $m_i^{}$ are the momentum, coordinates, and mass of the $i$-th atom, respectively.
Starting with a random initial momentum generated from the Maxwell-Boltzmann distribution,
 the equations of motion are solved for a discrete time step, $\Delta t_{\rm ML}^{}$,
 using a time-reversible and area-preserving algorithm (in the present study, the velocity-Verlet algorithm).
The ML potential surface $V_{\rm ML}^{}(\{\Vec{r} \},t)$ depends on time $t$ as it is trained on the fly,
 but it is kept constant within the time interval $\Delta t_{\rm T}^{}$, i.e.,
\begin{align}
V_{\rm ML}^{}(\{\Vec{r} \},t) &= V_{\rm ML}^n(\{\Vec{r} \}),  \: \: n \Delta t_{\rm T}^{} < t < (n+1) \Delta t_{\rm T}^{}, 
\end{align}
where $V_{\rm ML}^n$ is the ML potential for the $n$-th update.
$V_{\rm ML}^{}$ is trained every $\Delta t_{\rm T}^{} = n_{\rm DFT}^{} \Delta t_{\rm DFT}^{}$,
where $\Delta t_{\rm DFT}^{}$ $\equiv n_{\rm ML}^{} \Delta t_{\rm ML}^{}$ is the interval of acceptance/rejection in the Metropolis algorithm.
Thus $n_{\rm DFT}^{}$ is the number of times the DFT energy is computed for training,
and $n_{\rm ML}^{}$ is the number of steps in a trial move.
The acceptance probability for a trial move within the phase space from $\{\Vec{p},\Vec{r} \}$ to $\{\Vec{p}',\Vec{r}' \}$
is given by 
\begin{align}
&P_{\rm acc}^{}(\{\Vec{p},\Vec{r} \} \rightarrow \{\Vec{p}',\Vec{r}' \}) \nonumber \\
&= 
{\rm min} \left(1, e^{-\beta (H_{\rm DFT}^{}(\{\Vec{p}',\Vec{r}' \}) - H_{\rm DFT}^{}(\{\Vec{p},\Vec{r} \}))} \right), \label{eq:acc}
\end{align}
where $\beta = 1/T$ is the inverse of temperature and
\begin{align}
H_{\rm DFT}^{} &= \sum_i^N \frac{|\Vec{p}_i^{}|^2}{2 m_i^{}} + V_{\rm DFT}^{}(\{\Vec{r} \}).  \label{eq:hx}
\end{align}
is the Hamiltonian based on the DFT potential energy, $V_{\rm DFT}^{}(\{\Vec{r} \})$.
We note that the detailed balance condition is preserved exactly
 on the basis of the DFT (not ML) potential as long as
 the ML potential $V_{\rm ML}^{}(\{\Vec{r} \},t)$ does not change
 during a trial move.


According to Equations (\ref{eq:ri}) and (\ref{eq:acc}),
 ML force calculations are required $n_{\rm ML}^{}$ times
 while DFT energy calculations are required once.
SLHMC is computationally efficient when
 the former is less expensive than the latter,
 which is usually the case unless
 we set a huge value for $n_{\rm ML}^{}$.
In such a case, $\Delta t_{\rm ML}^{}$ could be
 chosen to be sufficiently small so as to
 conserve the ML energy within a trial move.
(This situation is different from conventional HMC
 where the step size should be large to break 
 energy conservation.)
When the ML energy is conserved within a trial move,
 the acceptance probability (\ref{eq:acc}) becomes 
\begin{align}
&P_{\rm acc}^{}(\{\Vec{p},\Vec{r} \} \rightarrow \{\Vec{p}',\Vec{r}' \}) \sim  
{\rm min} \left(1, e^{-\beta \Delta \Delta V} \right),  \label{eq:pacc}
\end{align}
where $\Delta \Delta V \equiv \Delta V(\{\Vec{r}' \}) - \Delta V(\{\Vec{r} \})$ and $\Delta V(\{\Vec{r} \}) \equiv V_{\rm DFT}^{}(\{\Vec{r} \})- V_{\rm ML}^{}(\{\Vec{r} \},t)$.
Therefore the accuracy of the ML potential influences the acceptance ratio, and thus the efficiency of the SLHMC method.
In practice, it is important that $n_{\rm ML}^{}$ is chosen
 carefully to give a decent acceptance ratio (more than circa 30\%).
\color{black}

In this paper, we use the Behler-Parrinello ANN potentials as ML potentials. 
We follow the standard ANN training protocol  to minimize the mean-square error ($\Delta V(\{ \Vec{r} \})^2$) \cite{Behler2007}.
The ANN variables are restarted from the last update ($n-1$) and optimized using all the DFT energy data up to the current update ($n$).
Training is possible using either only the accepted structures, or both the accepted and rejected structures (the latter is employed here).
Since the size of the training data set needed is small ($\sim 2000$ here), the computational cost of training the ANN potential is usually not dominant in SLHMC.  
As the number of training data increase with increasing $n$, one can skip the training steps after the ANN variables are optimized sufficiently by judging from the average acceptance ratio calculated in SLHMC.

\begin{figure}[t]
\begin{center}
     \begin{tabular}{p{ 1 \columnwidth}} 
      \resizebox{1 \columnwidth}{!}{\includegraphics{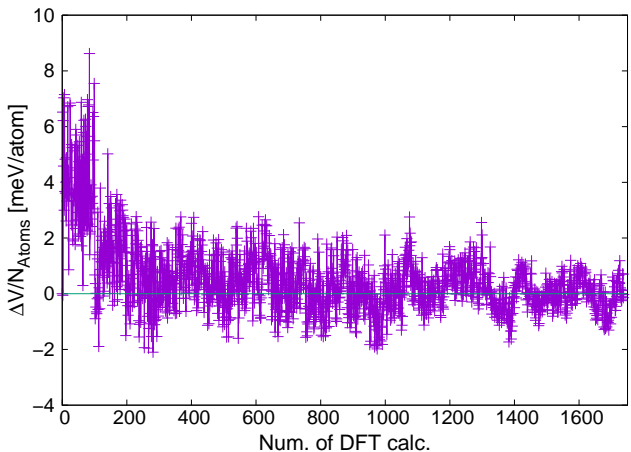}} 
    \end{tabular}
\end{center}
\caption{
\label{fig:diff}
Difference between the DFT and ANN energies of SiO$_2^{}$ in SLHMC. 
}
\end{figure}

\begin{figure}[t]
\begin{center}
     \begin{tabular}{p{ 1 \columnwidth}} 
  \resizebox{1 \columnwidth}{!}{\includegraphics{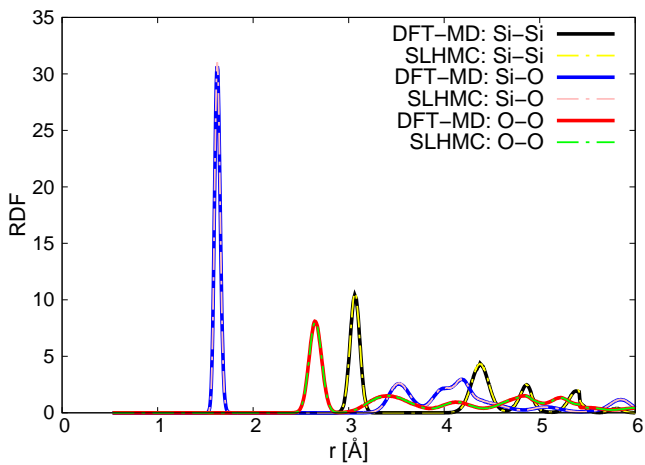}} 
    \end{tabular}
\end{center}
\caption{
\label{fig:rdf}(Color online) 
Radial distribution functions of SiO$_2^{}$ compared between DFT-MD and SLHMC. 
}
\end{figure}

\begin{figure}[th]
\begin{center}
     \begin{tabular}{p{ 1 \columnwidth}} 
   \resizebox{1 \columnwidth}{!}{\includegraphics{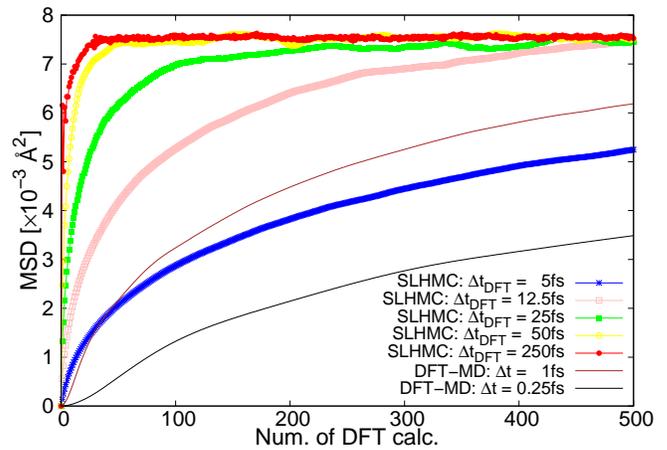}} 
    \end{tabular}
\end{center}
\caption{
\label{fig:msd}(Color online) Mean-squared displacement (MSD) of Si atoms for SiO$_2^{}$ calculated by the SLHMC at 300 K. 
The horizontal axis is the number of the DFT calculations. We set $\Delta t_{\rm ML} = 0.25$ fs in SLHMC. 
}
\end{figure}

\paragraph{Demonstration I: Thermodynamics of SiO$_2^{}$.}

The SLHMC method was implemented in the PIMD code \cite{PIMD}
 which has the access to ANN potentials and DFT calculations
 via the
 Atomic Energy Network (\ae net) \cite{ArtrithAENET}
 and Vienna Ab initio Simulation Packages (VASP) \cite{VASP}.
The first test was on the thermodynamics of the
 $\alpha$-quartz phase of SiO$_2^{}$ crystal \cite{Kimizuka2003}.
The SLHMC simulations were at 300 K
 in the canonical ensemble for a periodic system
 of 24 Si atoms and 48 O atoms.
The DFT calculations used
the Perdew-Burke-Ernzerhof (PBE) functional~\cite{PBE}.
The projected augmented wave (PAW) method~\cite{PAW} was employed, while
 the cut-off energy was 500 eV and the sampling points were gamma-point only. 
The ANN potentials were trained
 using the limited-memory Broyden-Fletcher-Goldfarb-Shanno
 (L-BFGS) method~\cite{Byrd1995,Zhu1997}. 
We adopt the Chebyshev basis set as a descriptor for
 atomic environments~\cite{Artrith2017},
 and the corresponding parameters are given
 in Ref.~\cite{note2}. 
The step size of ML, DFT and training were chosen
 to be $\Delta t_{\rm ML}^{} = 0.25$ fs, $\Delta t_{\rm DFT}^{} = 5$ fs, 
 and $\Delta t_{\rm T}^{} = 500$ fs, respectively.
The initial guess of the ANN potential was prepared
 and trained using a short DFT-MD trajectory of
 300 steps starting from the crystal structure.
 This could be done in other ways if
 the SLHMC run gets stuck at the initial step.

Figure \ref{fig:diff} shows that
 the difference between the ANN and DFT potentials
 quickly diminishes to about 1 meV/atom on the average
 as the SLHMC simulation proceeds
 and the ANN potential is trained on the fly.
This accuracy can be ascribed to the fact that
 the ANN potential has been trained in a confined
 configuration space corresponding
 to the exact ensemble at the DFT level.
Figure \ref{fig:rdf} shows that the radial distribution
 functions (RDFs) obtained from the SLHMC and DFT-MD
 simulations are identical, as they should be. 
 The RDFs in the SLHMC converge faster than those in the DFT-MD. 
Note that this is the case even when the ANN potentials
 are changed during the SLHMC simulation.
This demonstrates that the SLHMC method is able to
 gain statistics on thermodynamic properties
 while training the ANN potential at the same time.


Figure \ref{fig:msd} shows the results of
 the mean squared displacement (MSD)
 as a function of the number of DFT calculations,
 obtained with SLHMC and DFT-MD simulations.
The MSD is defined as
 $(1/N_{\rm Si}^{})
 \sum_{k=1}^{N_{\rm Si}^{}}
 ({\bf r}_k^\prime(t) - {\bf r}_k^\prime(0))^2$,
 where $N_{\rm Si}^{}$
 is the number of Si atoms and
 ${\bf r}_k^\prime(t)$ is the position of $k$-th atom
 relative to the center-of-mass of the system at time $t$.
These MSD curves,
 which should converge
 to a constant value for large $t$
 in solid states, indicate the length scale of
 the autocorrelation of atomic displacement.
As expected, the MSD converges faster as
 $\Delta t_{\rm DFT}^{}$ increases,
 since the ANN-MD trajectory is longer.
Assuming that the DFT calculations are the bottleneck
 of the SLHMC simulations, which was mostly the case
 in our computations, SLHMC simulations become
 more efficient as the MSD converges.
This is not only because the sampling of statistics
 becomes more efficient, but also because the training data
 becomes less correlated.
When $\Delta t_{\rm DFT}^{}$ is long enough to
 be uncorrelated in a single MC step,
 the efficiency of SLHMC should become
 proportional to the acceptance ratio,
 which is eventually reflected by
 the difference between the ANN and DFT potentials
 and thus the quality of the ANN.
In the present simulation,
 the acceptance ratio with the well-trained ANNs
 with $\Delta t_{\rm DFT}^{}=50$ fs was around 40\%,
 and the difference between the ANN and DFT potentials
 tended to 0.23 meV/atom on the average.

\begin{figure}[t]
\begin{center}
     \begin{tabular}{p{ 1 \columnwidth}} 
      \resizebox{1 \columnwidth}{!}{\includegraphics{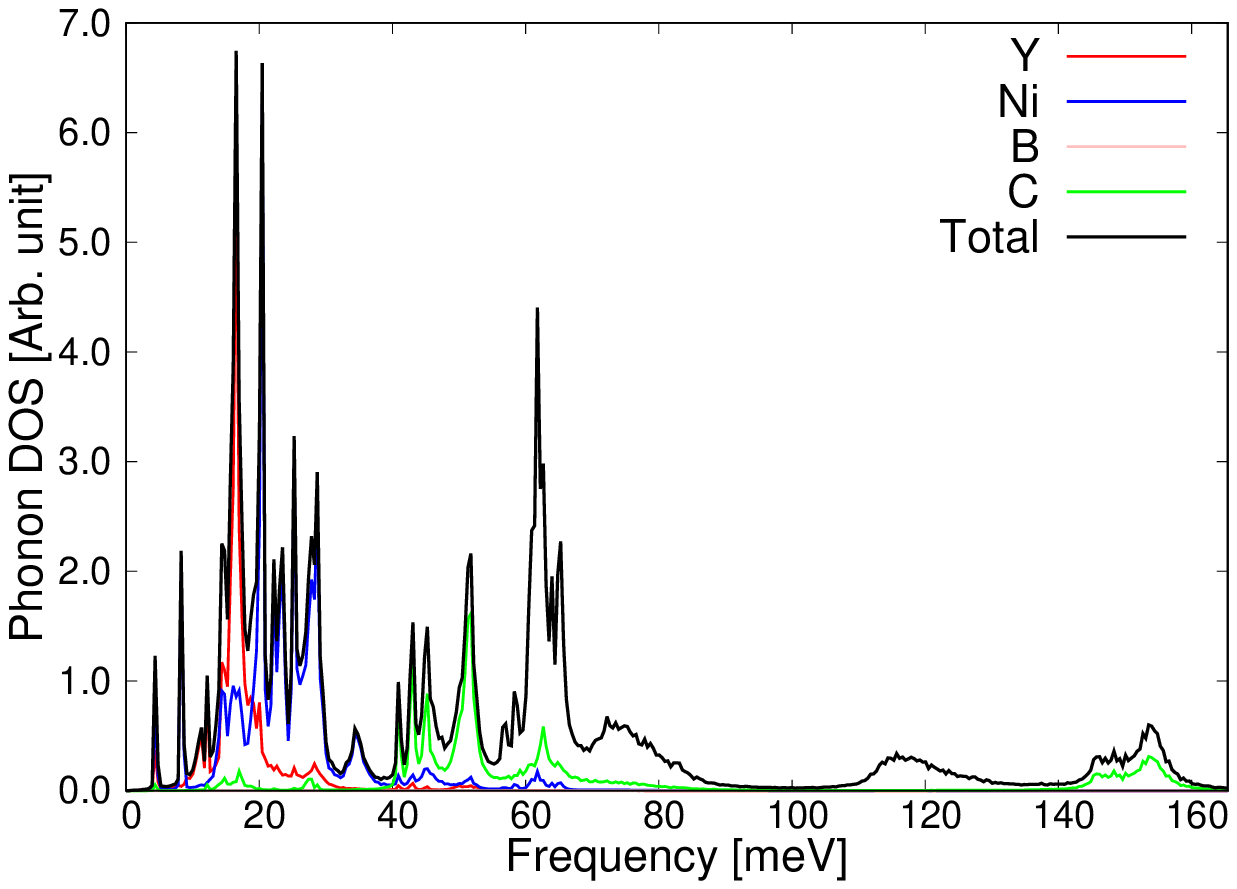}} 
      \resizebox{1 \columnwidth}{!}{\includegraphics{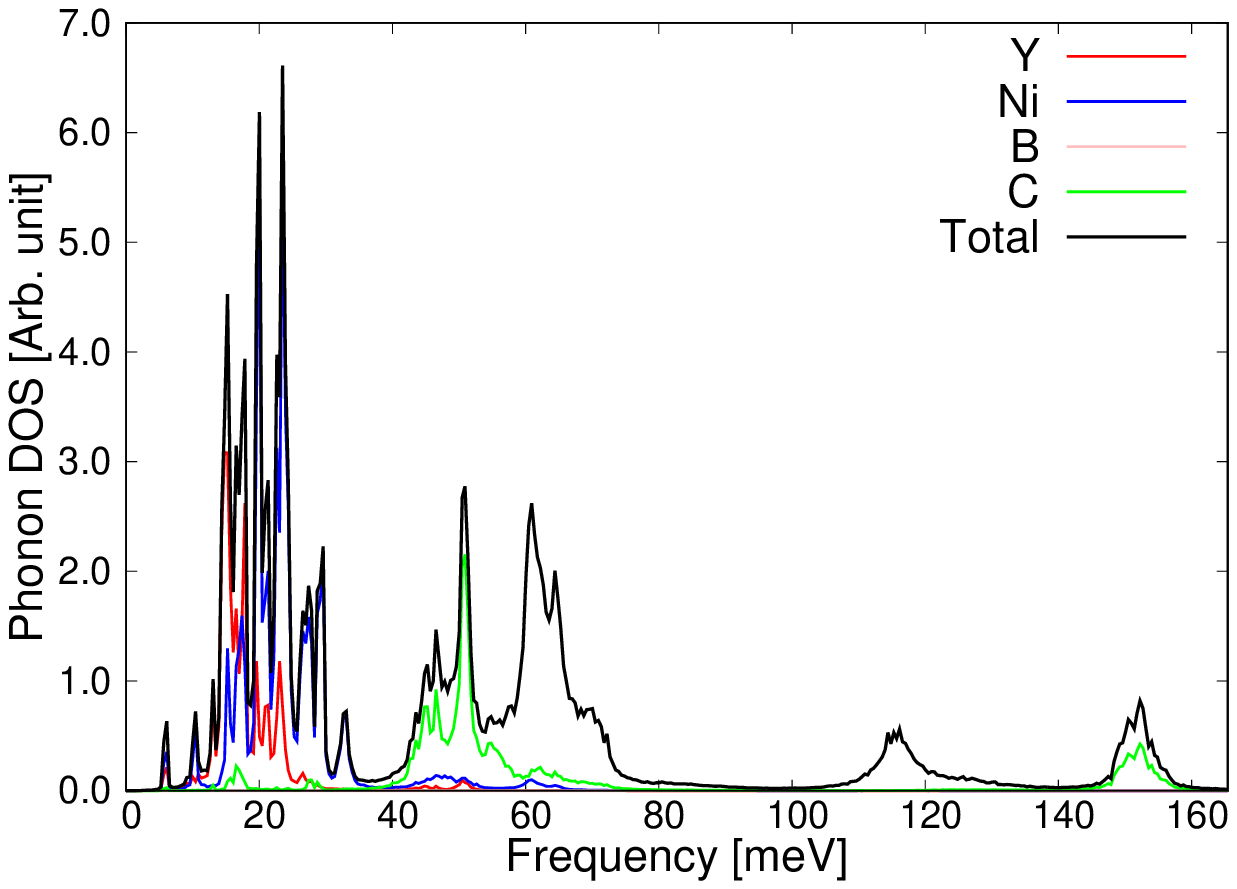}} 
    \end{tabular}
\end{center}
\caption{
\label{fig:fig3} (Color online) Phonon density of states in YNi$_2^{}$B$_2^{}$C calculated by ANN-MD with the ANN trained by the SLHMC. 
The temperatures are 60 K (upper panel) and 300 K (lower panel).
}
\end{figure}

\paragraph{Demonstration II: Dynamics of YNi$_2^{}$B$_2^{}$C.}

The second test was on the dynamics of
 an unconventional phonon-mediated
 superconductor YNi$_2^{}$B$_2^{}$C
 ($T_c^{} \sim 15$~K) \cite{Cava,NagaiYNi2B2C,Park,Izawa}.
Neutron scattering experiments have shown a
 strong temperature dependence
 of the phonon density of states (DOS)
 in this superconducting compound~\cite{Weber}.
The temperature dependence arises
 from anharmonic vibrations that are not
 taken into account in static DFT calculations 
 based on (quasi)-harmonic analysis
 at zero temperature \cite{LiuYNi2B2C,Weber}.
We show that a combination of
 the SLHMC and ANN-MD methods could be useful in this case.
Since the SLHMC method optimizes the ANN
 potential in the configuration space of
 a given ensemble,
 the accuracy of ANN potential is guaranteed
 in ensuing ANN-MD simulations as the trajectories 
 stay confined within
 this configuration space.
Therefore, once the ANN potential is optimized in SLHMC,
 the ANN-MD runs should be stable for long times,
 which may not necessarily be the case
 for complex systems with several elements
 (in the present case, four elements)
 for the methods proposed earlier.


The SLHMC and ANN-MD simulations were carried out for
 a supercell containing 16 Y atoms, 32 Ni atoms, 32 B atoms,
 and 16 C atoms.
The ANN and DFT simulations were setup in the same way as
 those in the previous section.
The SLHMC simulations were at 1000 K
 with the step sizes $\Delta t_{\rm ANN}^{} = 0.25$ fs,
 $\Delta t_{\rm DFT}^{} = 2.5$ fs, and
 $\Delta t_{\rm T}^{} = 250$ fs.
The difference of the trained ANN and DFT potentials
 was found to be about 0.4 meV/atom on average.
The DOS was computed via the Fourier transform of
 velocity autocorrelations~\cite{Goddard}
 from the ANN-MD simulations.
For statistical convergence,
 ten Newtonian trajectories were run independently
 for 100 ps with the step size
 $\Delta t_{\rm MD}^{} = 1$ fs.
These trajectories were restarted from the
 equilibrated configurations
 of NVT ensemble at 60 K and 300 K to reflect
 the temperature dependence~\cite{Perez} of the DOS.
As expected, all the ANN-MD trajectories
 were found to be stable for 100 ps.
\color{black}


As shown in Figure \ref{fig:fig3},
 the phonon density of states depends on temperature,
 which is consistent with the neutron
 scattering experiments~\cite{Weber}. 
This result confirms that anharmonic effects
 of phonons are important in this material.
The crystal structure of YNi$_2^{}$B$_2^{}$C is similar to that of high-$T_c$ cuprates and strongly anisotropic superconducting pairing has been suggested \cite{Park,Izawa,NagaiYNi2B2C}. 
Thus anharmonic effects of phonons appear to play a key role in this unconventional superconductor.


\paragraph{Conclusions.}

We proposed a new method called SLHMC
 to compute thermodynamic properties exactly
 based on DFT using an approximate ML potential that
 is trained on the fly to accelerate sampling.
The ML potential is
 optimized automatically by using a training data set of
 a given ensemble that is generated exactly.
The ML potential thus obtained can be used
 safely in ML-MD simulations to compute
 dynamic properties in thermal equilibrium,
 since the ML-MD simulations are stable for long times.
Proof-of-concept calculations were presented for
 the thermodynamics of SiO$_2^{}$ and
 the dynamics of YNi$_2^{}$B$_2^{}$C, which demonstrated the
 usefulness of SLHMC.


As can be expected from the acceptance probability
 in Eq. (\ref{eq:pacc}),
 the efficiency of SLHMC depends
 strongly on the balance between the system size
 and the quality of the ML potential.
Thus, it is recommended to keep the system
 reasonably small for SLHMC in practice.
However, like other ANN methods
 in the framework of the Behler-Parrinello approach,
 the ANN potentials
 are transferable to larger systems
 once they are well-trained by SLHMC.
Unlike the original HMC method,
 the small step size $\Delta t_{\rm ML}^{}$
 prevents deterioration of the efficiency of SLHMC.
The SLHMC approach established herein is not limited
 to solids, but it could be 
 applied generally to many kinds of systems such as molecular
 clusters and liquids.
In principle, this idea
 could be extended to other statistical ensembles,
 such as the isobaric ensemble and quantum ensembles
 via imaginary-time path integral theory~\cite{Tuckerman}.

\paragraph{Acknowledgment}
The calculations were performed on the supercomputing system SGI ICE X at the Japan Atomic Energy Agency.
This work was supported by JSPS KAKENHI Grant Numbers
 18K03552 (Y.N.), 18K05208 (M.S. and M.O.), 
 18H01693, 18H05519 and 16K05675 (M.S.).
 We thank Alex Malins for proofreading the manuscript.


\begin{thebibliography}{99}
\bibitem{note2}
The ANN fingerprint parameters were set as radial\_Rc = 8.0, radial\_N = 16, angular\_Rc = 6.5, and angular\_N = 4 for all the atomic elements.
We used two hidden layers with 15 units and the tanh-type activation function.
\end{thebibliography}
\end{document}